\documentclass{sig-alternate}
\usepackage{amsmath}
\usepackage{algorithm}
\usepackage{algorithmic}
\usepackage{multirow}
\usepackage{graphicx}
\usepackage{epsfig}

\begin{document}

\title{Hierarchical Bayesian Models with Factorization for Content-Based Recommendation}

\numberofauthors{1}
\author{
\alignauthor
Lanbo Zhang, Yi Zhang\\
       \affaddr{School of Engineering}\\
       \affaddr{UC Santa Cruz}\\
       \affaddr{Santa Cruz, CA, USA}\\
       \email{\{lanbo, yiz\}@soe.ucsc.edu}
}
\maketitle
\begin{abstract}

Most existing content-based filtering\footnote{``filtering'' and ``recommendation'' are interchangeable in this paper} approaches learn user profiles independently without capturing the similarity among users. Bayesian hierarchical models \cite{Zhang:Efficient} learn user profiles jointly and have the advantage of being able to borrow discriminative information from other users through a Bayesian prior. However, the standard Bayesian hierarchical models assume all user profiles are generated from the same prior. Considering the diversity of user interests, this assumption could be improved by introducing more flexibility. Besides, most existing content-based filtering approaches implicitly assume that each user profile corresponds to exactly one user interest and fail to capture a user's multiple interests (information needs).

In this paper\footnote{A short version of this paper is \cite{Zhang:2010:DFP:1871437.1871674}}, we present a flexible Bayesian hierarchical modeling approach to model both commonality and diversity among users as well as individual users' multiple interests. We propose two models each with different assumptions, and the proposed models are called Discriminative Factored Prior Models (DFPM). In our models, each user profile is modeled as a discriminative classifier with a factored model as its prior, and different factors contribute in different levels to each user profile. Compared with existing content-based filtering models, DFPM are interesting because they can 1) borrow discriminative criteria of other users while learning a particular user profile through the factored prior; 2) trade off well between diversity and commonality among users; and 3) handle the challenging classification situation where each class contains multiple concepts. The experimental results on a dataset collected from real users on digg.com show that our models significantly outperform the baseline models of L-2 regularized logistic regression and traditional Bayesian hierarchical model with logistic regression.

\end{abstract}




\section{Introduction}
Personalized recommendation has achieved great success in industrial applications, such as \textit{Amazon}, \textit{Netflix}, \textit{Google News}, etc. As an alternative to search engines, recommender systems actively present items (products, movies, news articles, books, images, etc.) that are likely of interest to the user.

There are two families of algorithms for personalized recommendation: collaborative filtering and content-based filtering \cite{Zhang:2013:CFS:2604906}. In this paper, we will focus on the content-based filtering. Existing content-based filtering research is largely influenced by the Filtering Track organized by TREC, where the task is to identify documents relevant to a specific topic from a document stream. There are two major approaches to handle this task. One is to use traditional IR retrieval models including Boolean model, traditional probabilistic models\cite{Robertson}, vector space model\cite{Rocchio}, language models\cite{Croft}, etc. These approaches are initially designed for ranking and need to be combined with threshold setting algorithms\cite{384012} when used for filtering. Another approach is to treat recommendation as a classification task, and thus many existing machine learning approaches such as Naive Bayes, Decision Tree, Logistic Regression, SVM, Neural Networks, etc. could be used. However, both the traditional IR models and the traditional machine learning models learn each user profile independently and do not make use of the commonality among users.

A major challenge for a recommender system is the cold-start problem\cite{564421}. The recommendation performance for new users (or infrequent users) is usually poor since the amount of training data from such users is small. This problem may badly hurt the development of a new recommender system since it means new users must endure the poor initial performance. There has been much research in machine learning fields on improving classification accuracy when only little training data is available. For example, semi-supervised learning is a way to use both unlabeled and labeled data to achieve this goal\cite{Zhu}; another approach is to introduce domain knowledge\cite{1148255}. However, these approaches are not specific for the recommendation task and thus may not fit the recommendation task well. Yu et al\cite{1009053} and Zhang et al\cite{Zhang:Efficient} introduced the Bayesian hierarchical modeling approach to \textbf{\textit{jointly}} learn user profiles for content-based filtering. Based on the fact that different users may have similar interests, the Bayesian hierarchical models assume that all user profiles are sampled from a common Gaussian prior. The Bayesian hierarchical modeling approach helps alleviate the cold-start problem since it is able to borrow discriminative information from other users through the common prior when learning a particular user profile, especially for those users with little training data. 



However, some users may have totally different interests, and requiring these users' profiles to follow the same prior distribution may negatively influence the learned profiles, thus we need to trade off better between the diversity and commonality of user profiles. Besides, almost all existing content-based filtering approaches cannot capture the multiple interests of a user. They implicitly assume each user profile only corresponds to a single interest, which does not fit the real scenarios in personalized recommendation, where a real user's interests may contain multiple concepts/topics. For example, a graduate student working on information retrieval may be interested in both IR research advancements and NBA news.




To better model the diversity and commonality of users and individual users' multiple interests, we propose a more flexible Bayesian hierarchical modeling approach for personalized content-based recommendation. The proposed models aim at borrowing discriminative information, modeling the diversity of users, and capturing the multiple interests of each user. We propose a parameter learning algorithm based on point estimation for the proposed models. We evaluate the proposed models and the parameter learning algorithm on a real recommendation dataset. The experimental results show that the proposed models can significantly outperform the state-of-the-art content-based filtering algorithms.

\section{Related work}
The central task of a recommender system is to find relevant items for individual users. The two basic families of recommendation approaches are collaborative filtering and content-based filtering.
\textbf{Collaborative filtering} doesn't need the content of an item and serves one user by leveraging information from other users with similar tastes and preferences. With its increasingly successful application in industry (e.g., Amazon, Netflix, Google News), collaborative filtering has attracted many researchers' attention during past several years. The memory-based \cite{Yu}\cite{Koren} and model-based \cite{963774}\cite{Luo:Flexible} approaches have been used in collaborative filtering task. \textbf{Content-based filtering} has been researched by the IR community for many years, especially in TREC 4-11 \cite{TREC1995FILTERING}\cite{TREC1996FILTERING}\cite{TREC1997FILTERING} \cite{TREC1998FILTERING}\cite{TREC1999FILTERING}\cite{TREC2000FILTERING}\cite{TREC2001FILTERING}\cite{TREC2002FILTERING}. Content-based filtering is the family of algorithms that make use of the characteristics of items and aim to understand which characteristics of items a particular user may like according to the user's history. In the scenario of content-based filtering, the recommender system maintains a profile/classifier\footnote{From the machine learning point of view, a user profile is a classifier} for each user, which is usually in the form of a number of features and their corresponding weights. The Rocchio algorithm\cite{Rocchio}, in which user profiles and documents are represented as term vectors, can gradually update user profiles when relevance feedback from users is available \cite{zhang2009ucsc,Xing:2011:BPR:2063576.2063866,Zhang:2010:IRB:1835449.1835511,Zhang:2011:FSD:2009916.2010003,Zhang:2012:SHS:2348283.2348306,Chen:2012:SBF:2396761.2398723}. Zhang et al proposed to combine Rocchio algorithm with logistic regression through a Bayesian prior \cite{1009052}. This algorithm outperforms both Rocchio algorihtm and logistic regression. Ault et al\cite{Tom} used kNN and Rocchio for information filtering. Cancedda et al\cite{Nicola}, Srikanth et al\cite{Srikanth} and Lewis\cite{David} used SVM for TREC filtering tasks. Stricker et al\cite{Stricker} used neural networks to handle the case of few relevant examples in filtering.

\begin{figure}[t]
\centering
\epsfig{file=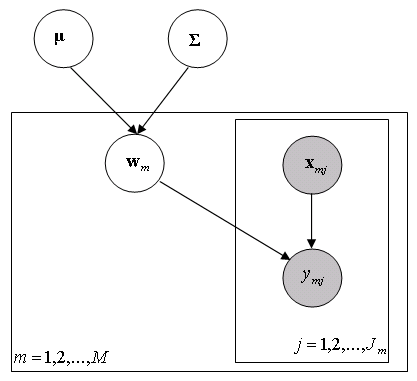, scale=0.7}
\caption{The Bayeisn Hierarchical Model (BHM). $\mathbf{\mu}$ is the mean vector of the Gaussian prior, and $\mathbf{\Sigma}$ is the covariance matrix which is usually assumed diagonal. The shaded nodes denote observed variables.}\label{fig:BHM}
\end{figure}

Among all content-based filtering approaches, the Bayesian hierarchical model has achieved the state-of-the-art performanes\cite{Zhang:Efficient,1009053}. It is quite related to our work, so we briefly describe it here. The most common Bayesian hierarchical model is shown in Figure \ref{fig:BHM} (we will use \textbf{BHM} to denote this model in the rest of this paper). There are $M$ users in total and each user has $J_m$ training examples. $\mathbf{w}_m$ is the profile of user $m$. All user profiles follow a Gaussian prior distribution with mean vector $\mathbf{\mu}$ and covariance matrix $\mathbf{\Sigma}$.

The other related work is the factorization-based topic models. For examples, Probabilistic Latent Semantic Analysis (PLSA)\cite{PLSA} and Latent Dirichlet Allocation (LDA)\cite{LDA} are two generative models that model documents as a mixture of latent topics. These models are very attractive theoretically. However, we are not aware of any empirical evidence of their effectiveness in competitive retrieval and filtering tasks. This is not surprising since most of the successful models in these tasks are discriminative models, such as Logistic Regression, SVM or Neural Networks, instead of generative models, such as Naive Bays. The discriminative factorization models in this paper are motivated by the theoretical attractiveness of topic models as well as the empirical success of discriminative models. Though our models also introduce latent factors, there are clear differences between our models and PLSA/LDA. Unlike PLSA and LDA, our models are discriminative models, and their goals are to learn user profiles which are in the form of discriminative functions. The entries of each factor in our models are not necessarily words, and could be any item features such as the time or authority of a document.

Another much related work is multi-task learning (aka transfer learning, learning to learn) in the machine learning community. Several methods have been proposed to learn related tasks together, focusing on capturing the commonality among the tasks through a shared representation or a shared prior. The Bayesian hierarchical modeling approach can be viewed as a multi-task learning framework based on a shared prior, and the models in \cite{Zhang:Efficient,1009053} and this paper are both special cases of this framework. Zhang et al\cite{Jian} proposed a probabilistic model to support a set of latent variable models for different multi-task learning scenarios. Their model is very similar to ours, however, the existing paper was focusing on multi-task learning and used Reuters document classification task for evaluation, while our work emphasizes on the application of personalized recommendation, which has very different characteristics and challenges.

\begin{figure}[t]
\centering
\epsfig{file=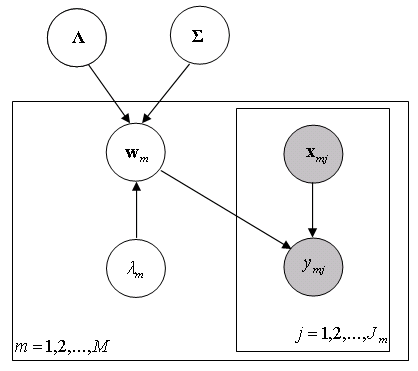, scale=0.7}
\caption{The Discriminative Factored Prior Model. The user profile $\textbf{w}_m$ follows a normal distribution with mean $\mathbf{\Lambda}\mathbf{\lambda}_m$ and variance $\mathbf{\Sigma}$. $\mathbf{\Lambda}$ is a $H$-column matrix with each column representing a hidden factor. Each entry of vector $\mathbf{\lambda}_m$ reflects how much each hidden factor contributes to the user profile.} \label{fig:dfpm}
\end{figure}

\section{Discriminative Factored\\ Prior Models (DFPM)}

The Discriminative Factored Prior Models (Figure \ref{fig:dfpm}) are motivated by the Bayesian hierarchical model in Figure \ref{fig:BHM}. $\mathbf{\Lambda}$ is a $K\times H$ matrix, and $\mathbf{\lambda_m}$ is a user-dependent vector with length $H$. The product of $\mathbf{\Lambda}$ and $\mathbf{\lambda}_m$ determines the prior mean of each user profile $\mathbf{w}_m$, and $\mathbf{\Sigma}$ is the prior covariance matrix. The assumption of DFPM is: there are a number of hidden factors that represent different concepts in the item feature space; users may be interested in one or several of these hidden factors in different levels. Each column of matrix $\mathbf{\Lambda}$, which is a $K$-dimendional vector, represents a specific hidden factor. $\mathbf{\lambda}_m$ tells how much each column of $\mathbf{\Lambda}$ contributes to the profile of user $m$.

To make our model description clear, we summarize the notations that will be used as follows: \\

\begin{itemize}
\item $H$: the total number of hidden factors.
\item $K$: the total number of item/document features.
\item $m=1,2,...,M$: the index of individual users. $M$ is the total number of users.
\item $j=1,2,...,J_m$: the index of training examples of user $m$. $J_m$ is the number of training examples for user $m$.
\item $\textbf{w}_m$: the profile of user $m$, which is a $K$-dimensional vector.
\item $\mathbf{\lambda}_m$: a user-related ($m$) $H$-dimensional vector that tells how much each hidden factor contributes to the user profile.
\item $<\textbf{x}_{mj},y_{mj}>$: the $j$-th training example of user $m$. $\textbf{x}_{mj}$ is a $K$-dimensional vector representing a training item, and $y_{mj}=1/0$ means this item is relevant/irrelevant to the user.
\item $\mathbf{\Lambda}$: a $K\times H$ matrix. Each column of $\Lambda$ represents a hidden factor.
\item $\mathbf{\Sigma}$: a $K\times K$ covariance matrix.
\end{itemize}

$\mathbf{\lambda}_m$ may follow two alternative distributions: Multinomial and Normal, and we will use \textbf{DFPM-Mult} and \textbf{DFPM-Norm} to denote these two models respectively. In DFPM-Mult, only one entry of $\mathbf{\lambda}_m$ is allowed to be 1 and all other entries be zero. This model clusters users into $H$ groups, and users of the same group share a common hidden factor as the prior. We want to point out that the common Bayesian hierarchical model (BHM) is actually a special case of DFPM-Mult. When $H=1$, DFPM-Mult is equivalent to BHM. In the case of DFPM-Norm, $\mathbf{\lambda}_m$ follows a Normal distribution with mean $\bf{0}$ and variance $b\bf{I}$. DFPM-Norm assumes that each user may be interested in multiple hidden factors, and each entry of $\mathbf{\lambda}_m$ reflects how much the corresponding hidden factor is related to the user's interests. DFPM-Norm is used to capture each user's multiple interests.

We assume each user profile $\textbf{w}_m$ is a random sample from a normal distribution with mean $\mathbf{\Lambda}\mathbf{\lambda}_m$ and variance $\mathbf{\Sigma}$. The label $y_{m_j}$ of a training item $\mathbf{x}_{m_j}$ is $y = f(\mathbf{x}_{m_j};\mathbf{w}_m)$, where $f$ could be many existing regression or classification models. We will take the logistic regression as an example to demonstrate how our models could be used for recommendation task. Let $\bf{I}$ be an identity matrix, $a,b,c$ be constant parameters, we summarize the discriminative factored prior models with logistic regression as follows:

\begin{itemize}
\item Each column of hidden factor matrix $\mathbf{\Lambda}$ follows a normal distribution: $N(\bf{0}, a\bf{I})$. $\mathbf{\Sigma}$ follows an Inverse Wishart distribution $\bf{W}^{-1}(\bf{I},c)$
\item The user vector $\mathbf{\lambda}_m$ may follow two alternative distributions: Normal or Multinomial. In the case of DFPM-Norm, $\mathbf{\lambda}_m \sim N(\bf{0}, b\bf{I})$; and in the case of DFPM-Mult, $\mathbf{\lambda}_m \sim Multinomial(\frac{1}{H}, \frac{1}{H},...,\frac{1}{H})$
\item The user profile $\textbf{w}_m$ follows a normal distribution: $\textbf{w}_m \sim N(\mathbf{\Lambda}\mathbf{\lambda}_m, \mathbf{\Sigma})$
\item Given a user profile $\textbf{w}_m$ and an item feature vector $\textbf{x}_{mj}$, its label is sampled from: $$y_{mj} \sim Bernoulli(\frac{1}{1+exp(-\textbf{w}_m^T \textbf{x}_{mj})})$$
\end{itemize}

Let $\theta=(\mathbf{\Lambda}, \mathbf{\Sigma}, \mathbf{\lambda}_m, \textbf{w}_m)$ be the parameters of our models that need to be estimated. The joint likelihood of all variables and the observed data $D=\left\{<\textbf{x}_{mj},y_{mj}>\right\}$ is:
\begin{align}
\begin{split}
& P(D,\theta)   = \\
& P(\mathbf{\Lambda})P(\mathbf{\Sigma})\prod_{m=1}^{M}{P(\mathbf{\lambda}_m)P(\textbf{w}_m|\mathbf{\lambda}_m, \mathbf{\Lambda}, \mathbf{\Sigma})\prod_{j=1}^{J_m}{P(y_{mj}|\textbf{x}_{mj}, \textbf{w}_m)}}
\end{split}
\label{likelihood}
\end{align}

\section{Parameter Estimation}
The empirical Bayes method\cite{EBM} is often used when learning a complicated Bayesian hierarchical model like DFPM. However, the learning algorithm based on Empirical Bayes will be very complicated since there are many hidden variables ($\mathbf{\Lambda}$, $\mathbf{\Sigma}$, $\mathbf{\lambda}$, $\textbf{w}_m$) entangled in our models. Besides, the number of features and the number of users involved in the recommendation task are usually huge. Thus the empirical Bayes method is too computationally expensive to be used here. As an alternative, we present a simplified learning algorithm based on point estimation. This algorithm calculates a single value (the best guess) for each unknown variable.

\subsection{The Learning Algorithm} \label{sec:la}
To simplify the learning process, we assume the variance of $\textbf{w}_m$ is given, and as a result, $\mathbf{\Sigma}$ is removed from our models. Based on Equation \ref{likelihood}, the maximum likelihood estimation of the parameters can be found based on Equation \ref{equ:obj}.\footnote{In this section, we use $y_{mj}=-1$ to represent the label of a negative training example} We introduce the constants $c_1,c_2,c_3$ to replace the variance $\mathbf{\Sigma}, b, a$ respectively, and their model effects are equivalent.

\begin{algorithm}[t]
\caption{\label{alg:dfpm}{\bf: The Parameter Learning Algorithm}}
\begin{algorithmic}
\STATE 1: Initialize $\mathbf{\Lambda}^0$, $i \leftarrow 0$
\STATE 2: Repeat:
\STATE 3: ~~For each user $m$, compute $\textbf{w}_m^i$, $\mathbf{\lambda}_m^i$ based on $\mathbf{\Lambda}^i$ by solving Equation \ref{equ:wl}
\STATE 4: ~~Compute $\mathbf{\Lambda}^{i+1}$ based on $\textbf{w}_m^i$, $\mathbf{\lambda}_m^i$ of each user according to Equation \ref{equ:sLR}
\STATE 5: ~~$i \leftarrow i+1$
\STATE 6: Until convergence
\STATE 7: Return $\mathbf{\Lambda}, \mathbf{\lambda}_m, \textbf{w}_m$
\end{algorithmic}
\end{algorithm}

\begin{algorithm}[t]
\caption{\label{alg:dfpm-norm}{\bf: The Step 3 of Model DFPM-Norm in Algorithm \ref{alg:dfpm}}}
\begin{algorithmic}
\STATE 1: Initialize $\mathbf{\lambda}_m^{i,0}$, $h \leftarrow 0$
\STATE 2: Repeat:
\STATE 3:  ~~Compute $\textbf{w}_m^{i,h}$ based on $\mathbf{\lambda}_m^{i,h}$ by solving Equation \ref{equ:w}
\STATE 4:  ~~Compute $\mathbf{\lambda}_m^{i,h+1}$ based on $\textbf{w}_m^{i,h}$ by Equation \ref{equ:sl}
\STATE 5:  ~~$h \leftarrow h+1$
\STATE 6: Until convergence
\STATE 7: Return $\textbf{w}_m^i, \mathbf{\lambda}_m^i$
\end{algorithmic}
\end{algorithm}

\begin{algorithm}[t]
\caption{\label{alg:dfpm-mult}{\bf: The Greedy Algorithm for Step 3 of Model DFPM-Mult in Algorithm \ref{alg:dfpm}}}
\begin{algorithmic}
\STATE 1: $c \leftarrow \arg\min_{c}{\sum_{j=1}^{J_m}\log\left[1+\exp(-y_{mj}(\mathbf{\Lambda}_c^i)^T \textbf{x}_{mj})\right]}$
\STATE 2: Let the $c$-th entry of $\mathbf{\lambda}_m^i$ be 1 and all else be 0
\STATE 3: Compute $\textbf{w}_m^i$ based on $\mathbf{\lambda}_m^i$ by solving Equation \ref{equ:w} using gradient based method
\STATE 4: Return $\textbf{w}_m^i, \mathbf{\lambda}_m^i$
\end{algorithmic}
Note: $\mathbf{\Lambda}_c^i$
 is the $c$-th column of matrix $\mathbf{\Lambda}^i$
\end{algorithm}

\begin{align}
\label{equ:obj}
\begin{split}
& \left(\hat{\mathbf{\Lambda}},\hat{\mathbf{\lambda}}_m,\hat{\textbf{w}}_m\right)= \\
&\arg\min_{\mathbf{\Lambda},\mathbf{\lambda}_m,\mathbf{w}_m}\left\{\sum_{m=1}^{M}\left\{\sum_{j=1}^{J_m}\log\left[1+\exp(-y_{mj}\textbf{w}_m^T \textbf{x}_{mj})\right] \right.\right. \\
& +
\left.\left. \vphantom{\sum_{}^{}} c_1\left\|\textbf{w}_m-\mathbf{\Lambda}\mathbf{\lambda}_m\right\|^2+c_2\left\|\mathbf{\lambda}_m\right\|^2\right\}+c_3\left\|\mathbf{\Lambda}\right\|^2\right\}
\end{split}
\end{align}

To solve Equation \ref{equ:obj}, we first initialize $\mathbf{\Lambda}^0$, then update the parameters by following the iteration steps A and B below. $i$ is the iteration number. This process is also summarized in Algorithm \ref{alg:dfpm}.

\begin{itemize}
\item \textbf{Step A: find $\textbf{w}_m^i$ and $\mathbf{\lambda}_m^i$ for each user based on $\mathbf{\Lambda}^i$.}
\end{itemize}

Given $\mathbf{\Lambda}^i$, the user profiles are independent from each other. Thus we can focus on each individual user and solve the following subproblem for each user:
\begin{equation}
\label{equ:wl}
\begin{split}
(\textbf{w}_m^i, \mathbf{\lambda}_m^i) = & \arg\min_{\textbf{w}_m, \mathbf{\lambda}_m}\left\{\sum_{j=1}^{J_m}\log\left[1+\exp(-y_{mj}\textbf{w}_m^T \textbf{x}_{mj})\right] \right.\\
& \left. \vphantom{\sum_{}^{}} + c_1\left\|\textbf{w}_m-\mathbf{\Lambda}^i \mathbf{\lambda}_m\right\|^2+c_2\left\|\mathbf{\lambda}_m\right\|^2\right\}
\end{split}
\end{equation}

To solve Equation \ref{equ:wl}, in the case of DFPM-Norm, we first initialize $\mathbf{\lambda}_m^{i,0}$, then follow the iteration steps 1 and 2 below to find the solution. $h$ is the iteration number. This process is also summarized in Algorithm \ref{alg:dfpm-norm}.

\begin{itemize}
\item \textit{Step 1: given $\mathbf{\lambda}_m^{i,h}$, calculate $\textbf{w}_m^{i,h}$}
\end{itemize}

\begin{equation}
\label{equ:w}
\begin{split}
\textbf{w}_m^{i,h} = & \arg\min_{\textbf{w}_m}\left\{\sum_{j=1}^{J_m}\log\left[1+\exp(-y_{mj}\textbf{w}_m^T \textbf{x}_{mj})\right] \right.\\
& \left. \vphantom{\sum_{}^{}} +c_1\left\|\textbf{w}_m-\mathbf{\Lambda}^i\mathbf{\lambda}_m^{i,h}\right\|^2\right\}
\end{split}
\end{equation}

There is no close form solution for Equation \ref{equ:w}, so we choose to use gradient based method such as conjugate gradient decent\cite{CGD} to solve it. Let $L$ be the loss function in Equation \ref{equ:w}, we can derive the following derivative for $\textbf{w}_{m}$:
\begin{equation}
\label{equ:d}
\frac{\partial L}{\partial \textbf{w}_m} = \sum_{j=1}^{J_m}{\frac{-y_{mj}\textbf{x}_{mj}}{1+\exp(y_{mj}\textbf{w}_m^T \textbf{x}_{mj})}} + 2c_1(\textbf{w}_m-\mathbf{\Lambda}^i\mathbf{\lambda}_m^{i,h})
\end{equation}

\begin{itemize}
\item \textit{Step 2: caluclate $\mathbf{\lambda}_m^{i,h+1}$ based on $\textbf{w}_m^{i,h}$}
\end{itemize}

\begin{equation}
\label{equ:l}
\begin{split}
\mathbf{\lambda}_m^{i,h+1} = \arg\min_{\mathbf{\lambda}_m}\left\{ c_1\left\|\textbf{w}_m^{i,h}-\mathbf{\Lambda}^i\mathbf{\lambda}_m\right\|^2+c_2\left\|\mathbf{\lambda}_m\right\|^2 \right\}
\end{split}
\end{equation}

By taking derivative of $\mathbf{\lambda}_m$ on the loss function of Equation \ref{equ:l} and let it equal 0, we can get the following close form solution for Equation \ref{equ:l}:
\begin{equation}
\label{equ:sl}
\begin{split}
\mathbf{\lambda}_m^{i,h+1} = \left( {\mathbf{\Lambda}^i}^T\mathbf{\Lambda}^i + c_2\textbf{I} \right)^{-1}{\mathbf{\Lambda}^i}^T\textbf{w}_m^{i,h}
\end{split}
\end{equation}

In the case of DFPM-Mult, there are $H$ alternative values for $\mathbf{\lambda}_m$. A straightforward approach is to try each possible value of $\mathbf{\lambda}_m$, and use a gradient search method to solve Equation \ref{equ:w}. The $\mathbf{\lambda}_m$ and $\textbf{w}_m$ that lead to the minimum loss function value could be found finally. However, this approach is very time-consuming in real applications, especially when $H$ is huge. Instead, we use a greedy algorithm (Algorithm \ref{alg:dfpm-mult}) to approximately solve Equation \ref{equ:wl}. The basic idea is to set $\textbf{w}_m=\mathbf{\Lambda}^i\mathbf{\lambda}_m$ and find $\mathbf{\lambda}_m$ that minimizes the loss function in Equation \ref{equ:wl}. With the $\mathbf{\lambda}_m$ found, we then use gradient search to find the best $\textbf{w}_m$ based on Equation \ref{equ:w}. This method is less time-consuming since the gradient search is only run once.

\begin{itemize}
\item \textbf{Step B: find $\mathbf{\Lambda}^{i+1}$ based on $\textbf{w}_m^i$ and $\mathbf{\lambda}_m^i$ of all users.}
\end{itemize}
\begin{equation}
\label{equ:L}
\begin{split}
\mathbf{\Lambda}^{i+1}=\arg\min_{\mathbf{\Lambda}}\left\{ \sum_{m=1}^M c_1\left\|\textbf{w}_m^i-\mathbf{\Lambda}\mathbf{\lambda}_m^i\right\|^2 + c_3\left\|\mathbf{\Lambda}\right\|^2 \right\}
\end{split}
\end{equation}

Different rows of $\mathbf{\Lambda}$ are independent with each other, thus Equation \ref{equ:L} could be solved row by row. Let $\mathbf{\Lambda}_r^{i+1}$ be the $r$-th row vector of $\mathbf{\Lambda}^{i+1}$, and $\textbf{w}_{m,r}^i$ be the $r$-th element of $\textbf{w}_m^i$, then for $r=1,2,...,K$,
\begin{equation}
\label{equ:LR}
\begin{split}
\mathbf{\Lambda}_r^{i+1}=\arg\min_{\mathbf{\Lambda}_r}\left\{ \sum_{m=1}^M c_1\left(\textbf{w}_{m,r}^i-\mathbf{\Lambda}_r\mathbf{\lambda}_m^i\right)^2 + c_3\left\|\mathbf{\Lambda}_r\right\|^2 \right\}
\end{split}
\end{equation}

By taking derivative of $\mathbf{\Lambda}_r$ on the loss function of Equation \ref{equ:LR}, we get the following solution for Equation \ref{equ:LR}:
\begin{equation}
\label{equ:sLR}
\begin{split}
\mathbf{\Lambda}_r^{i+1} = \left( \left( \sum_{m=1}^{M}{\mathbf{\lambda}_m^i{\mathbf{\lambda}_m^i}^T + \frac{c_3}{c_1}\textbf{I}} \right)^{-1}\sum_{m=1}^{M}{\textbf{w}_{m,r}^i\mathbf{\lambda}_m^i} \right)^T
\end{split}
\end{equation}

\subsection{Time Complexity}
\label{sec:time}
In each iteration, the major computation work comprises the following three parts:
\begin{itemize}
\item The gradient search process involves the computation of the loss function in Equation \ref{equ:w} and the derivative in Equation \ref{equ:d}. This takes time $O(Mln(\bar J+H)K)$, where $M$ is the number of users, $l$ is the total iteration number in algorithm \ref{alg:dfpm-norm} (for DFPM-Norm) or 1 (for DFPM-Mult), $n$ is the average steps of the chosen gradient search algorithm, $\bar J$ is the average number of training examples for each user, $H$ is the number of hidden factors, and $K$ is the number of features. This part takes up most of the running time.
\item For DFPM-Norm, the computation of Equation \ref{equ:sl}, whose time complexity\footnote{We assume the time complexity of the matrix inversion algorithm is in $O(H^3)$.} is $O(Ml(H^3+HK))$. For DFPM-Mult, this step is not included.
\item The computation of Equation \ref{equ:sLR}. For DFPM-Norm, the time complexity is $O(H^3+MHK)$, and for DFPM-Mult, since $\mathbf{\lambda}_m$ has only one non-zero entry, the matrix that needs to be inverted is diagonal, thus the time complexity is $O(MHK)$.
\end{itemize}

We conclude that the time complexity of our learning algorithm is linear to: the number of users, the number of training examples, and the number of features. Besides, DFPM-Norm and DFPM-Mult are respectively cubic and linear to the number of hidden factors.

\section{Experimental Methodology}

\subsection{The Dataset}
To evaluate the proposed modeling approach, we collected a data set from Digg.com\cite{Digg}. Digg.com is a website for people to share web content including news, images, and videos. Users can digg items they are interested in to promote the items' ranking. Each item dugg by a user is considered a positive data point (a relevant document) for the user. We collected the complete digg history of news articles of more than 15,000 users. The detailed statistics of our dataset is shown in Table \ref{tab:digg}.

Since Digg.com only has user digg history available on its website, we couldn't get those articles users read but didn't digg. In other words, we don't have real negative examples. To address this problem, we randomly choose equal number\footnote{We didn't make the positive and negative classes unbalanced since the unbalanced problem is not the focus of this paper.} of articles which are not dugg by a user as the negative examples for this user. Considering the large percentage of user undugg articles, we expect most of the articles sampled in this way are irrelevant to this user's interests.


\begin{table}
\caption{Statistics of the crawled Digg data}\label{tab:digg}
\begin{center}
\begin{tabular}{|c|c|}
\hline
Number of users & 15,162 \\
\hline
Number of news articles & 91,088 \\
\hline
Total number of diggs & 3,809,196 \\
\hline
Average number of diggs per user & 251 \\
\hline
\end{tabular}
\end{center}
\end{table}

\subsection{Experimental Details}

We randomly divide each user's data (including both positive and negative examples) into three parts: training (80\%), validation (10\%), and test (10\%). The validation data is used to tune the parameters of both our models ($H,c_1,c_2,c_3$) and the baseline models. 
We use the words as features. Both the stop words and rare words (occurring in less than 50 articles) are removed from the feature set. As a result, there are 35,865 features. When calculating the feature values, we use the TF*IDF scoring method.

\textbf{Precision}, \textbf{Recall}, and \textbf{Macro-F1} are used as the evaluation measures. The conjugate gradient decent algorithm implemented in \cite{Macopt} is used in our implementation.

\subsection{Experimental Goals}
Our experiments are designed to answer the following questions:
\begin{itemize}
\item How is the performance of our models compared with the state-of-the-art algorithms?
\item Can our models learn meaningful hidden factors, and how does the number of hidden factors ($H$) influence the performance?
\item Which assumption about $\mathbf{\lambda}_m$ works better, Multinomial or Normal distribution?
\item How is the efficiency (running time) of our models?
\end{itemize}

To answer the first question, we compare our models with the L-2 regularized logistic regression (\textbf{L2LR}) and the Bayesian hierarchical model (Figure \ref{fig:BHM}) with logistic regression (\textbf{BHLR}) implemented in \cite{HMBBR}. By comparing our models with L2LR, which learns each user profile independently, we can tell whether our models can borrow useful discriminative information from other users. By comparing our models with BHLR, which is a state-of-the-art algorithm for content-based filtering \cite{Zhang:Efficient}, we want to evaluate whether our models can perform better by introducing a factored prior.

To answer the second question, the performances with different numbers of hidden factors will be compared. We will analyze the learned hidden factors and see whether the dominating features of each hidden factor look reasonable.

To answer the third question, we compare the two models corresponding to two assumptions of $\mathbf{\lambda}_m$: the multinomial (\textbf{DFPM-Mult}) and normal (\textbf{DFPM-Norm}) distributions.

To answer the fourth question, we will record the running time of our learning algorithms with different numbers of users ($M$) and different numbers of hidden factors ($H$). We will see the overall running time of our algorithms and how the running time increases with the increase of the user number and the hidden factor number.

\section{Experimental Results}
\subsection{Overall Performances}
\begin{figure*}
\centering
\epsfig{file=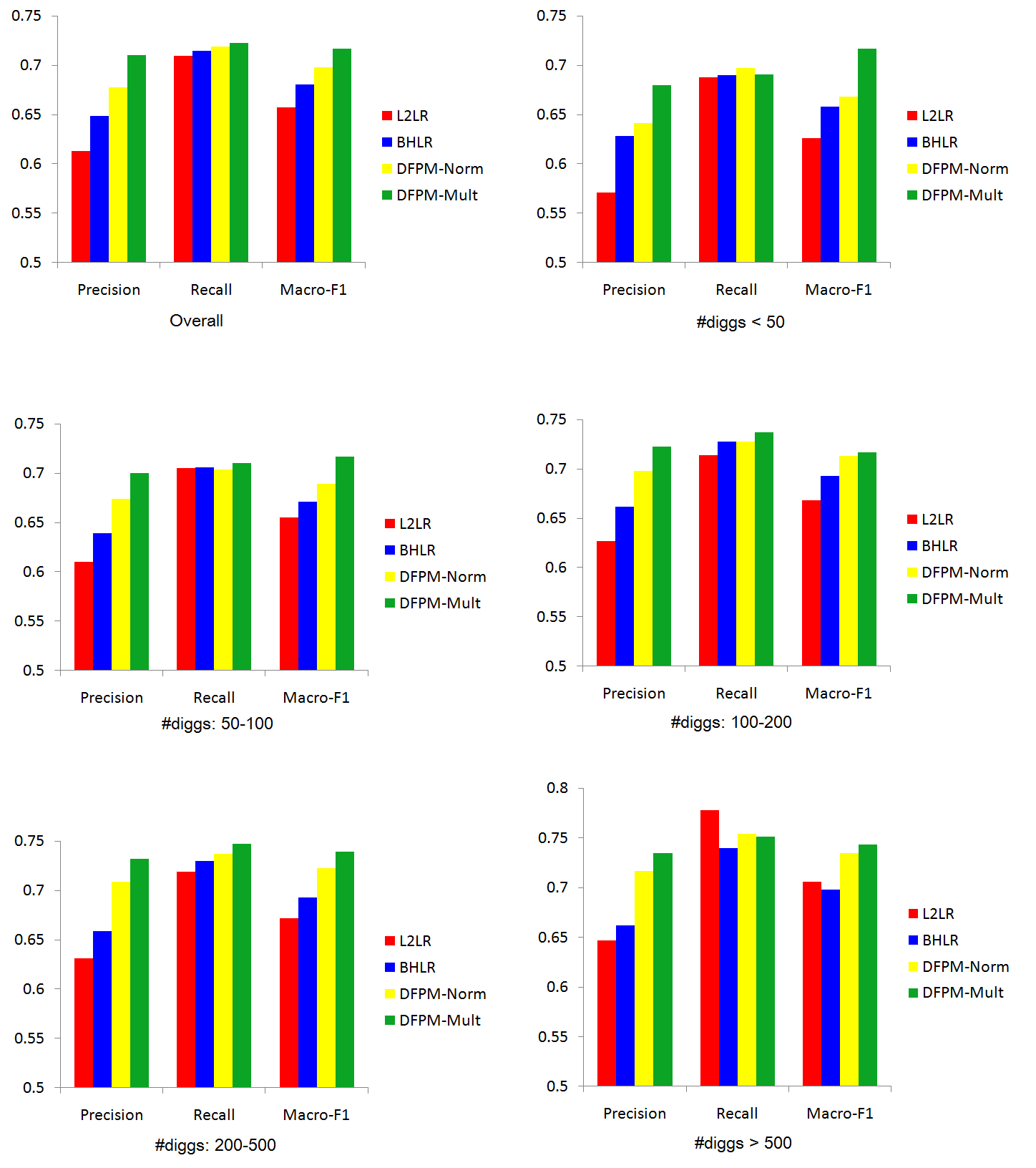, scale=0.45}
\caption{Performance comparison of different models. L2LR: L-2 regularized logistic regression. BHLR: Bayesian hierarchical logistic regression model. DFPM-Norm and DFPM-Mult are our models. The top-left graph shows the overall performance on all users, and the rest on five user groups with different \#diggs: less than 50, 50-100, 100-200, 200-500, greater than 500 respectively. According to our statistical test (t-test) results, our models \textit{significantly} outperform the baseline methods.}\label{fig:pc}
\end{figure*}

The top-left graph in Figure \ref{fig:pc} shows the overall performances of four algorithms. Both of our models (DFPM-Norm and DFPM-Mult) are statistically \textit{\textbf{significantly}} better than the baselines in terms of Precision and Macro-F1 (based on t-test). The improvement on precision is very significant. This is very encouraging since Precision is a more important factor in most personalized recommender/filtering systems. To see whether our algorithms are helpful for both hard users (users with little training data) and easy users (users with much training data), we divide the users into five groups according to their numbers of diggs (less than 50, 50-100, 100-200, 200-500, greater than 500 respectively) to see whether the performances for all kinds of users have been improved. The rest graphs in Figure \ref{fig:pc} show the results on these user groups. We find our models outperform the baselines for all five user groups.

\begin{figure}
\centering
\epsfig{file=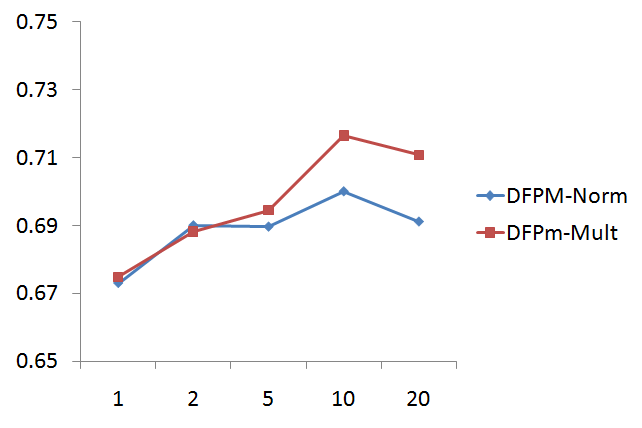, scale=0.35}
\caption{Performances (Macro-F1) of our models at different $H$ (number of hidden factors)}\label{fig:H}
\end{figure}

\subsection{DFPM v.s. L2LR}
Figure \ref{fig:pc} shows that our models \textbf{\textit{significantly}} outperform the L-2 regularized logistic regression, which learns each user profile independently. This demonstrates that our models successfully borrow discriminative information from other users by learning user profiles jointly.

\subsection{DFPM v.s. BHLR}
Figure \ref{fig:pc} shows that our models \textbf{\textit{significantly}} outperform the Bayesian hierarchical logistic regression model (BHLR). We find that BHLR already significantly outperforms L2LR, which indicates BHLR successfully borrows information from other users. Encouragingly, our models can further improve the performances. This demonstrates that our models can learn more accurate user profiles by introducing a factored prior.

Why our models can outperform BHLR? Not all users have similar interests, and it's not always a good idea to assume that all user profiles are generated from the same Gaussian distribution. Our models have less strong model assumptions and use the variable $\mathbf{\lambda}_m$ to model the diversity of users, and thus have the advantage of only borrowing information from \textbf{\textit{similar}} users. In particular, users with similar interests share a common hidden factor as the prior in the DFPM-Mult model.

\subsection{The Hidden Factors}
Figure \ref{fig:H} shows how the number of hidden factors influences the performance. Remember BHLR is a special case ($H=1$) of our model DFPM-Mult. As $H$ increases from 1 to 10, the performance keeps on improving and reaches the optimal value at $H=10$.  If we consider users with similar interests as a cluster, our model DFPM-Mult can effectively identify the underlying user clusters. To better understand the DFPM-Mult model, we list the most popular (dugg by most users) news articles of each user cluster in Table \ref{tab:5m}. It seems that the articles in each cluster somewhat represent similar topics. For example, the first cluster (Politics), the third (Apple products), and the fifth (Software). We also list the top 20 features (words) in some learned factors in Table \ref{tab:topwords}. We observe that most of the words represent the concept of each hidden factor well.

\begin{table*}
\caption{Top 5 most popular news articles of each user cluster (by model DFPM-Mult). For space limit, only the first five user clusters are shown.}\label{tab:5m}
\begin{center}
\begin{tabular}{|l|l|l|}
\hline
\textbf{Cluster NO.} & \textbf{\#users in the cluster} & \textbf{Top 5 news articles} \\
\hline
\multirow{5}{2cm}{1} & \multirow{5}{3cm}{1235} & Digg\_This\_If\_You\_Voted\_For\_Obama\_2 \\
&& Congress\_Passes\_Historic\_Health\_Care\_Reform \\
&& Barack\_Obama\_wins\_the\_2008\_Presidential\_Election \\
&& Barack\_Obama\_Officialy\_Becomes\_44th\_American\_President \\
&& We\_had\_eight\_years\_of\_Bush\_and\_Cheney\_Now\_you\_get\_mad\_2 \\
\hline
\multirow{5}{2cm}{2} & \multirow{5}{3cm}{255} & Guns\_Tumors\_And\_The\_Limits\_Of\_The\_Human\_Eye \\
&& High\_Tech\_Sport\_Suits\_May\_Give\_Skaters\_Gold\_Medal\_Edge \\
&& Lawmaker\_Wants\_to\_Force\_Kids\_to\_Wear\_Cups\_While\_Bike\_Riding \\
&& New\_Charging\_Method\_Could\_Slash\_Battery\_Recharge\_Times \\
&& Deep\_sea\_pigs\_used\_to\_investigate\_dead\_zones \\
\hline
\multirow{5}{2cm}{3} & \multirow{5}{3cm}{874} & Digg\_The\_Digg\_iFrame\_Toolbar\_is\_Dead\_Unbanning\_Domains \\
&& 8\_Things\_That\_Suck\_About\_the\_iPad \\
&& Google\_to\_offer\_ultra\_high\_speed\_broadband\_in\_US \\
&& Apple\_holding\_iPhone\_OS\_4\_event\_April\_8th \\
&& Curious\_user\_opened\_up\_his\_brand\_new\_iPad\_to\_see\_PIC \\
\hline
\multirow{5}{2cm}{4} & \multirow{5}{3cm}{1079} & Callers\_flood\_ehealthinsurance\_Where\_s\_My\_Free\_ObamaCare \\
&& Teen\_Sues\_Mom\_for\_Hacking\_Facebook\_Account \\
&& Childhood\_Deafness\_Gene\_Uncovered \\
&& Japan\_s\_whale\_meat\_obsession \\
&& Skater\_Jumps\_Fails\_Gets\_Eaten\_by\_Shed \\
\hline
\multirow{5}{2cm}{5} & \multirow{5}{3cm}{538} & Ubuntu\_9\_10\_Almost\_Perfect \\
&& Ubuntu\_dumps\_the\_brown\_introduces\_new\_theme\_and\_branding \\
&& VLC\_1\_0\_0\_Released \\
&& Linux\_Users\_Should\_Say\_Goodbye\_Apple \\
&& 10\_skills\_developers\_will\_need\_in\_the\_next\_five\_years \\
\hline
\end{tabular}
\end{center}
\end{table*}

\begin{table}
\caption{Top words in some factors (by model DFPM-Mult).}\label{tab:topwords}
\begin{center}
\begin{tabular}{|l|l|l|l|l|}
\hline
obama & scientist & smartphon & linux\\
presid & relationship & tablet & mozilla\\
jed & mlm & chrome & chrome\\
palin & exercis & android & diggtv\\
beck & geograph & dropbox & broadband\\
marijuana & treatment & feb & anonym\\
mccain & coach & broadband & techradar\\
yellow & copenhagen & diggtv & lifehack\\
barack & foreclosur & dialogg & dialogg\\
religi & orbit & chines & interfac\\
tortur & seti & tc & firefox\\
ko & npr & webcast & server\\
gop & emiss & idg & databas\\
pot & apprais & ie6 & ie6\\
limbaugh & climat & anon & hacker\\
congress & techniqu & bing & odomzig\\
sarah & tablet & vulner & laptop\\
stewart & attract & hothardwar & pcworld\\
lewison & treehugg & infograph & rekcahefil\\
bailout & flight & aol & css\\
\hline
\end{tabular}
\end{center}
\end{table}

\subsection{$\mathbf{\lambda}_m$: Multinomial v.s. Normal}
Figure \ref{fig:pc} shows that DFPM-Mult performs better than DFPM-Norm. This is somewhat surprising. Initially, we expected that DFPM-Norm should perform better than DFPM-Mult since the Normal assumption of $\mathbf{\lambda}_m$ can capture the multiple interests of individual users. There are several possible reasons for this finding. First, it's possible that the flexibility of $\mathbf{\lambda}_m$ makes the learning process more complicated. Second, the flexibility of $\mathbf{\lambda}_m$ may curtail the information borrowed from other users so that the commonality of similar users is not captured well. We are planning to investigate the reasons in more details in the future work.

\subsection{The Running Time}
\begin{figure*}
\centering
\epsfig{file=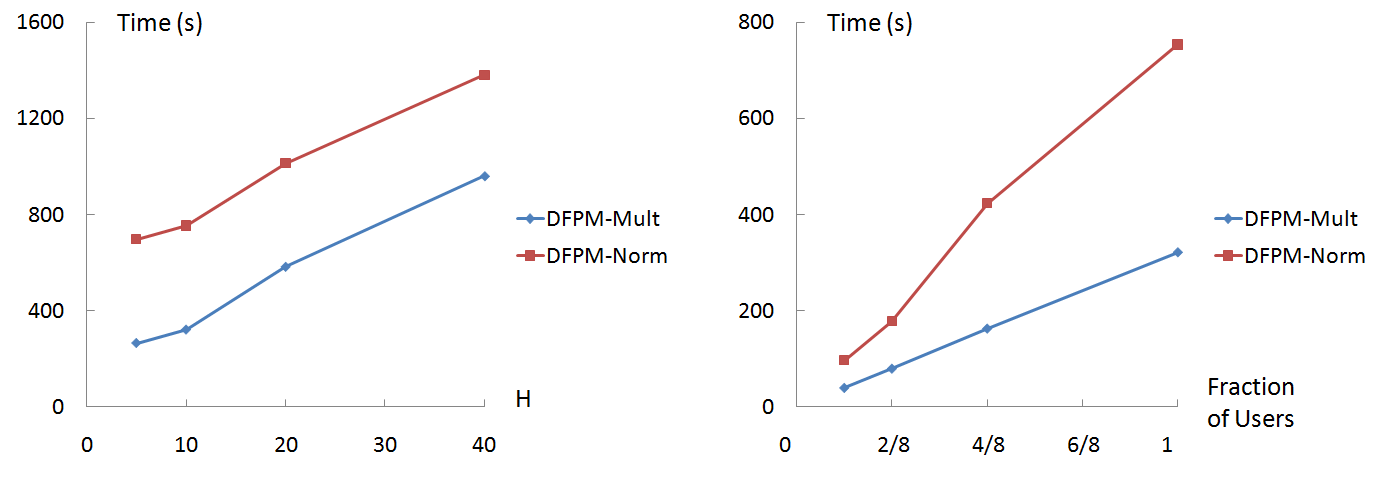, scale=0.35}
\caption{The running time (per iteration) of our learning algorithms. The left graph shows the running time (seconds) with different numbers of hidden factors($H$). The right graph shows the running time with different fractions of users (15,162 users in total). }\label{fig:time}
\end{figure*}

Figure \ref{fig:time} shows the running performance of our learning algorithms. We report the running time (seconds) per iteration for both DFPM-Mult and DFPM-Norm with different numbers of hidden factors (left) and different fractions of users (right). The experiment was run on an Intel Xeon CPU (E5420@2.5GHz) with 16GB memory. The results in Figure \ref{fig:time} agree with the time complexity analysis in Section \ref{sec:time}: 1) the time complexity of the learning algorithm for DFPM-Mult is linear to the number of users ($M$) and the number of hidden factors ($H$), and that for DFPM-Norm is linear to the number of users\footnote{We didn't find the time complexity of algorithm for DFPM-Norm is cubic to $H$, since when $H$ is small, the running time is dominated by the gradient search steps in section \ref{sec:la}}; 2) DFPM-Mult is more efficient since it doesn't include the inner iteration process in our learning algorithm (see section \ref{sec:la}). Our algorithms converge reasonably fast and all the results reported in Figure \ref{fig:pc} are obtained within 20 iterations.

\section{Conclusions}
In this paper, we present a more flexible Bayesian hierarchical modeling approach for personalized content-based recommendation. Particularly, we propose two Discriminative Factored Prior Models and the corresponding parameter learning algorithms. Our modeling approach aims at: 1) learning user profiles jointly and borrowing discriminative criteria from other users by modeling the commonality of users through a shared factorized prior; 2) modeling the diversity of users and only borrowing information from \textbf{\textit{similar}} users (DFPM-Mult); and 3) capturing multiple interests of individual users (DFPM-Norm).

We evaluate our models on a dataset collected from real web users on Digg.com\cite{Digg}, and compare them with two much related baseline algorithms. The experimental results demonstrate that:
\begin{itemize}
\item Our models significantly improve the recommendation performance, especially for users with little training data. Thus they can help alleviate the cold-start problem. 
\item It's helpful to introduce a factorized prior. Particularly, the DFPM-Mult model learns more accurate user profiles since it can effectively cluster users with similar interests and has the advantage of only borrowing discriminative criteria from similar users while learning a particular user profile. 
\item The time complexity of our parameter learning algorithms is linear to the number of users. Thus our algorithms are applicable in real recommender systems.
\item The multinomial assumption of $\mathbf{\lambda}_m$ seems to work better than the normal assumption.
\end{itemize}


In the future work, more research is needed to analyze the normal assumption of $\mathbf{\lambda}_m$, since this model captures each user's multiple interests and thus offers some advantages over the DFPM-Mult model. One possible approach is to try a Dirichlet prior instead of a normal prior for $\mathbf{\lambda}_m$. Besides, we will also evaluate our models on more datasets, using explicit user feedback on irrelevant items. Our models can also be modified to fit the personalized recommendation task better, for example, to capture user interest drift by adding temporal variables.


\bibliographystyle{abbrv}
\bibliography{cikm10}

\end{document}